\documentclass[twocolumn,showpacs,preprintnumbers,amsmath,amssymb]{revtex4}

\usepackage{pstricks}
\usepackage{graphicx}
\usepackage{dcolumn}
\usepackage{bm}

\begin{document}

\def\F{{\bf F}}
\def\A{{\bf A}}
\def\J{{\bf J}}
\def\af{{\bf \alpha}}
\def\beqn{\begin{eqnarray}}
\def\eeqn{\end{eqnarray}}

\def\dspace{\baselineskip = .30in}
\def\beq{\begin{equation}}
\def\non{\nonumber}
\def\bwi{\begin{widetext}}
\def\ewi{\end{widetext}}
\def\pl{\partial}
\def\na{\nabla}
\def\al{\alpha}
\def\bt{\beta}
\def\eeq{\end{equation}}
\def\Ga{\Gamma}
\def\ga{\gamma}
\def\de{\delta}
\def\De{\Delta}
\def\da{\dagger}
\def\ka{\kappa}
\def\si{\sigma}
\def\Si{\Sigma}
\def\te{\theta}
\def\La{\Lambda}
\def\lam{\lambda}
\def\Om{\Omega}
\def\om{\omega}
\def\ep{\epsilon}
\def\sq{\sqrt}
\def\sqg{\sqrt{G}}
\def\sp{\supset}
\def\sb{\subset}
\def\l{\left (}
\def\r{\right )}
\def\lq{\left [}
\def\rq{\right ]}
\def\fr{\frac}
\def\la{\label}
\def\hs{\hspace}
\def\vs{\vspace}
\def\inf{\infty}
\def\ran{\rangle}
\def\lan{\langle}
\def\ov{\overline}
\def\tl{\tilde}
\def\tm{\times}
\def\lrar{\leftrightarrow}


\preprint{CERN-PH-TH-2006-195,~
OSU-HEP-06-8}

\title{Proton Stability In Supersymmetric $SU(5)$}

\author{Zurab Tavartkiladze}%
 \email{zurab.tavartkiladze@okstate.edu}
\affiliation{%
\it Physics Department, Theory Division, CERN, CH-1211 Geneva 23, Switzerland \\
Department of Physics, Oklahoma State University, Stillwater, OK 74078, USA
}%

\date{October 29, 2006}

\begin{abstract}

Within supersymmetric $SU(5)$ GUT we suggest
mechanisms for suppression of baryon number violating dimension five and six operators.
The mechanism is based on the idea of split multiplets (i.e. quarks and leptons are not
coming from a single GUT state) which is realized by an extension with additional
vector-like matter. The construction naturally avoids wrong asymptotic 
relation $\hat{M}_D=\hat{M}_E$. Thus, the long standing problems of the minimal SUSY
$SU(5)$ GUT can be resolved.

In a particular example of flavor structure and with additional ${\cal U}(1)\tm {\cal Z}_{3N}$
symmetry we demonstrate how the split multiplet mechanism works out. Namely, the considered model is compatible
with successful gauge coupling unification and realistic fermion mass pattern. The nucleon decay rates
are relatively suppressed and can be well compatible with current experimental bounds.

\end{abstract}

\pacs{12.10.Dm, 12.10.Kt, 12.15.Ff}
\maketitle

\section{Introduction}

Baryon number violation is one of the predictions of Grand Unified Theories (GUT). In SUSY GUTs, usually 
dimension five ($d=5$) operator induced proton decay dominates \cite{Sakai:1981pk}. The sources for the 
latter are heavy color triplets' couplings   with ordinary matter supermultiplets. These couplings usually 
originate from the operators responsible for quark and lepton masses. Therefore, the observed Yukawa couplings 
and the baryon number violating operators may be closely related and this is the reason that it is not easy to 
satisfy the present experimental
bound $\tau^{\rm exp}(p\to K\nu )\stackrel{>}{_\sim }6.7\cdot 10^{32}$~years \cite{ft1} on proton life time 
\cite{Yao:2006px}. On the other hand, it is not trivial to build realistic fermion mass pattern within GUTs.
Therefore, the task is two fold: 1) within considered scenario care must be exercised to get realistic fermion 
masses and mixings, and 2) within the same framework the baryon number violating processes must be suppressed 
up to the required level. These are two main problems and for their resolution numerous mechanisms and specific 
examples have been suggested
\cite{mechs, Berezhiani:1998hg, Shafi:1999rm, Shafi:1999tn, Bajc:2002bv, Nath:2006ut}.
It is a curious fact that the split multiplet mechanism, for suppressing the baryon number violation, more 
or less has been  ignored
(see however \cite{Shafi:1999rm, Shafi:1999tn}). Let us note that this mechanism is naturally realized
within extra dimensional constructions \cite{Kawamura:2000ev, Dermisek:2001hp}. This is, most likely, the reason that within 
four dimensional constructions there were not many attempts to realize and apply this possibility. However, once 
the multiplet splitting is achieved (i.e quarks and leptons come from different
GUT states), the baryon number can be conserved up to the needed level \cite{Dermisek:2001hp}.
In this paper we suggest mechanisms for natural quark-lepton splitting within four dimensional SUSY $SU(5)$ \cite{ft2}.
We show that apart from suppressing the baryon number violation this splitting enables one to build realistic 
fermion mass pattern. 
 The discussion of the mechanism and it's needed ingredients are presented in the next section.
In section \ref{sec:d6suppr} we show how $d=6$ nucleon decay can be suppressed.
In section \ref{sec:realSU5}, for demonstrative purposes, we present particular example in which split multiplet 
mechanism is realized. It utilizes an 
additional ${\cal U}(1)\tm {\cal Z}_{3N}$ symmetry which plays crucial role for adequate suppression of all unwanted
baryon number violating couplings including Planck scale suppressed operators. An assumption on a particular flavor structure 
and simple minded SUSY spectrum near$\sim 1$~TeV is made. These give perturbative gauge coupling unification and 
realistic fermion masses and mixings. At the same time, nucleon's decay rate is compatible with current
experimental bounds.

\section{Suppression of $d=5$ Baryon Number Violation}\label{sec:d5suppr}

In the minimal SUSY $SU(5)$ (MSSU5) GUT the matter sector consists of the
$(10+\bar 5)$-plets per generation with the following decomposition under
$SU(3)_c\tm SU(2)_L\tm U(1)_Y$:

$$
10=q(3, 2)_{-1}+u^c(\bar 3, 1)_4+e^c(1, 1)_{-6}~,
$$
\beq
\bar 5=d^c(\bar 3, 1)_{-2}+l(1, 2)_3~,
\la{dec10-5}
\eeq
where subscripts stand for the hypercharges in
$1/\sq{60}$ units [$Y=\fr{1}{\sq{60}}{\rm Diag}\l 2, 2, 2, -3, -3\r $].
The pair of scalar superfields $H(5)+\bar H(\bar 5)$ has the following composition:
$$
H(5)=h_u(1, 2)_{-3}+T(3, 1)_2~,
$$
\beq
\bar H(\bar 5)=h_d(1, 2)_3+\bar T(\bar 3, 1)_{-2}~,
\la{decH}
\eeq
where $h_u, h_d$ denote the MSSM higgs doublet superfields, and $T, \bar T$
are their colored GUT partners.
The renormalizable operators $10\cdot 10H$ and $10\cdot \bar 5\bar H$ (the family indices
are suppressed), together with ordinary Yukawa superpotential couplings,
generate matter-$T, \bar T$ interactions:
$$
\lam 10\cdot 10H=\lam \l qu^ch_u+qqT+e^cu^cT\r ~,
$$
\beq
\lam' 10\cdot \bar 5\bar H=
\lam' \l qd^ch_d+e^clh_d+ql\bar T+u^cd^c\bar T\r ~.
\la{hTcoupl}
\eeq
Integration of $T, \bar T$ states (with mass $M_T$) generates  $d=5$
operators
\beq
\fr{\lam \lam'}{M_T}\l qqql\r_F~,~~~~~~~\fr{\lam \lam'}{M_T}\l u^cu^cd^ce^c\r_F~,
\la{d5ops}
\eeq
which induce the nucleon decay. Current experimental bound on nucleon lifetime
requires $\lam \lam'\stackrel{<}{_\sim }10^{-9}$ (for $M_T\sim 10^{16}$~GeV and all soft SUSY breaking terms$\sim $TeV). On the other hand, in
MSSU5 $\lam \lam'$ is directly related to the quark and lepton Yukawa
couplings and is typically$\sim 10^{-6}/\sin 2\bt $. 
 This would lead to unacceptably fast proton decay. Note that couplings in (\ref{hTcoupl})
also lead to the wrong asymptotic mass relations $m_{\mu }=m_s$, $m_e/m_{\mu }=m_d/m_s$ at GUT scale.
Thus, some modification should be done anyway in order to improve this situation.
It is desirable to have a mechanism which simultaneously solve both - fermion mass problem and
baryon number violation. Note that modification of either Yukawa sector \cite{Berezhiani:1998hg, Bajc:2002bv},
or sparticle spectrum \cite{Shafi:1999vm}, or increase of GUT scale \cite{propos} can improve the situation with baryon number violation. 
However, besides colored higgsino mediated $d=5$ operators (\ref{d5ops}), there exist Planck scale suppressed
baryon number violating couplings which in SUSY $SU(5)$ have forms
\beq
\fr{\lam_{\rm Pl}}{M_{\rm Pl}}\l 10\cdot 10\cdot 10\cdot \bar 5\r_F
\to \fr{\lam_{\rm Pl}}{M_{\rm Pl}}\l qqql+u^cu^cd^ce^c\r_F ~.
\la{Planckd5ops}
\eeq
This also mediates proton decay and in order to satisfy experimental bound one should arrange for 
appropriate couplings $\lam_{\rm Pl}\stackrel{<}{_\sim }10^{-7}$. Couplings $\lam_{\rm Pl}$ are 
completely independent from the Yukawa sector and therefore their smallness need separate
explanation, because at SUSY $SU(5)$ level there is no symmetry argument for their suppression.

Below we present mechanism, different from existing ones,
which within SUSY $SU(5)$ GUT suppress (eliminate) baryon number violation and can solve the fermion mass problem.

\subsection{Suppressing $qqT$ and eliminating $e^cu^cT$ operators}\label{sec:qqTsuppr}

{}In order to demonstrate how the split multiplet mechanism works,  we start considerations
with one family only. The
generalization to three families is straightforward and will be discussed
later on. We extend the matter sector with vector like states in $15$
and $\bar{15}$ representations of $SU(5)$. In terms of $SU(3)_c\tm SU(2)_L\tm U(1)_Y$ they 
decompose as
\beq
15=q(3, 2)_{-1}+S(6, 1)_4+\De (1, 3)_{-6}~,
\la{dec-15}
\eeq
and conjugate transformations for the fragments of $\bar{15}=(\bar q, \bar S, \bar{\De })$.
The state $q$($\equiv q_{15}$) from $15$-plet
has transformation properties of the left handed quark doublet. The remaining
$S$ and $\De $ states have 'exotic' quantum numbers. This feature of
$15$-plet can be used for the suppression of proton decay
\cite{Shafi:1999rm}. With suitable couplings we can arrange that the light
left handed quark doublet mainly comes from  $15$-plet. Consider the
superpotential couplings
\beq
10\Si \bar{15}+M_{15}15\cdot \bar{15}~,
\la{10-15}
\eeq
where $\Si $ is an adjoint $24$-plet scalar superfield used for the breaking
$SU(5)\to SU(3)_c\tm SU(2)_L\tm U(1)_Y$.
Substituting in (\ref{10-15}) the GUT VEV $\lan \Si \ran \equiv M_G$
with $M_{15}\ll \lan \Si \ran$, we see that $q_{10}$ decouples by forming
the massive state with $\bar q_{\bar{15}}$. Namely, for the light $q$
and heavy $q_h$ states we have
$$
q\simeq q_{15}~,~~~q_h\simeq q_{10}+\fr{M_{15}}{M_G}q_{15}~~=>~~
$$
\beq
15\supset q~,~~~~~10\supset \ep q ~~~~{\rm with}~~~~\ep \equiv \fr{M_{15}}{M_G}~.
\la{comp-q}
\eeq
The states $u^c$ and $e^c$ (from $10$-plet) and fragments $(S, \De ), (\bar S, \bar{\De })$ (from
$15, \bar{15}$) are not affected with this procedure. Therefore,
\beq
u^c~,e^c\subset 10~,
\la{comp-ue}
\eeq
and masses of the decoupled states are given by
\beq
M(q_{10}, \bar q_{\bar{15}})\simeq M_G~,~~~~M_S=M_{\De }=M_{15}~.
\la{mas1}
\eeq
Now it is clear that the up quark mass will be generated through the Yukawa coupling
of $15$-plet with $10$. Since $15$-plet is the two index symmetric
representation of $SU(5)$, $\Si $ should participate in this coupling.
Namely,
$$
Y\fr{\Si }{M_*}15\cdot 10 H\to Y_U\l qu^ch_u+\ep qqT\r ~,~~~
$$
\beq
{\rm with}~~~~~~Y_U=\fr{\lan \Si \ran }{M_*}Y~.
\la{up-qqT}
\eeq
We see that the  term $qqT$ is suppressed by factor $\ep $ in comparison to
the  up type quark Yukawa coupling.
This occurred thanks to the splitting of the $q$-states living in
 $15$ and $10$ plet superfields respectively.
Note that no $e^cu^cT$ coupling arises from (\ref{up-qqT}). The
coupling $10\cdot 10H$  is not needed at all and can be suppressed
or completely eliminated in concrete scenario (discussed in sect. \ref{sec:realSU5}).

The scale $M_*$ in (\ref{up-qqT}) is a cut off and one expects that it is much larger than the
GUT scale $M_*\gg \lan \Si \ran $ (in most conservative approach
$M_*\sim M_{\rm Pl}\simeq 2.4\cdot 10^{18}$~GeV - the reduced Planck mass). Thus, we can use this
type of coupling for first two light families (i.e. for generation of up and charm quark masses).
For the top quark mass we need to have unsuppressed Yukawa coupling.
If we do not apply this mechanism of $qqT$ coupling suppression for the third
generation, the top Yukawa can be due to the coupling $10_310_3H$.
However, the same coupling also generates unsuppressed $q_3q_3T$ term. This
would give sizable contribution to the nucleon decay \cite{Hisano:1992jj, Berezhiani:1998hg}
through the mixings with light families.
Thus, for suppressing $q_3q_3T$ and generating the top Yukawa coupling at
renormalizable level we suggest slight modification by introducing additional
$10'+\bar{10}'$ states and the couplings
\beq
10\Si \bar{15}+15\Si \bar{10}'+M_{15}15\cdot \bar{15}+
M_{10}10'\cdot \bar{10}'
\la{10'-10-15}
\eeq
From these terms we can write down the mass matrices for appropriate
fragments
\beq
\begin{array}{ccc}
 & {\begin{array}{cc}
\hs{-0.2cm}\bar q_{\bar{10}'} \hspace{0.8cm} & \hspace{0.5cm}\bar q_{\bar{15}}
\end{array}}\\ \vspace{1mm}

\begin{array}{c}
 q_{10'}\vs{0.1cm}\\ q_{10}\vs{0.1cm} \\ q_{15}
 \end{array}\!\!\!\!\!\hs{-0.2cm} &{\left(\begin{array}{ccc}

\hs{0.6mm} M_{10}&\hs{0.3cm} 0
\\
\vs{-0.3cm}
\\
 \hs{0.6mm}0& \hs{0.3cm}M_G
 \\
 \vs{-0.3cm}
 \\
 \hs{0.6mm}M_G\hs{0.1cm}&\hs{0.3cm}M_{15}

\end{array}\right)},~
\end{array}
\begin{array}{cc}
 & {\begin{array}{c}
\hs{-0.2cm}\bar u^c_{\bar{10}'}
\end{array}}\\ \vspace{1mm}

\begin{array}{c}
u^c_{10}\vs{0.1cm}\\ u^c_{10'}
 \end{array}\!\!\!\!\!\hs{-0.02cm} &{\left(\hs{-0.15cm}\begin{array}{ccc}
0
\\
\vs{-0.3cm}
\\
M_{10}

\end{array}\hs{-0.15cm}\right)}~,
\end{array}
\begin{array}{cc}
 & {\begin{array}{c}
\hs{-0.2cm}\bar e^c_{\bar{10}'}
\end{array}}\\ \vspace{1mm}

\begin{array}{c}
e^c_{10}\vs{0.1cm}\\ e^c_{10'}
 \end{array}\!\!\!\!\! &{\left(\hs{-0.15cm}\begin{array}{ccc}
0
\\
\vs{-0.3cm}
\\
M_{10}

\end{array}\hs{-0.15cm}\right)}~.
\end{array}  \!\!
\label{compon10-15}
\end{equation}
The masses of the fragments $S$ and $\De $ are still given by (\ref{mas1}).
With $M_{15}\ll M_G$, $M_{10}\stackrel{<}{_\sim }M_G$ the masses of remaining decoupled states
are
$$
M(q_{10}, \bar q_{\bar{15}})\sim M(q_{15}, \bar q_{\bar{10}'})\simeq M_G~,
$$
\beq
M(u^c_{10'}, \bar u^c_{\bar{10}'})\simeq M(e^c_{10'}, \bar e^c_{\bar{10}'})
\simeq M_{10}~,
\la{mas2}
\eeq
and distribution of light $q$, $u^c$ and $e^c$ fragments will be as follows
\beq
10'\supset q ~,~~10\supset u^c~,e^c, \ep'q~,~~{\rm with}~~
\ep'\equiv \fr{M_{10}M_{15}}{M_G^2}~.
\la{comp-que}
\eeq
We will identify the $q$ state from $10'$ and $u^c$ from $10$ with the third generation matter.
Therefore, the up quark mass is generated through the coupling
$10'\cdot 10H$, while the $qqT$ coupling will be suppressed.
In more detail, taking into account (\ref{comp-que}) we will have
\beq
Y_U10'\cdot 10H\to Y_U\l qu^ch_u+\ep' qqT\r ~.
\la{up-qqT-1}
\eeq
Note that $e^cu^cT$ coupling is still not generated from (\ref{up-qqT-1}).

With these simple mechanisms we will be able to suppress
$d=5$ proton decay up to the needed level.
If for $i$-th generation ($i=1,2,3$) the suppression factor of the
corresponding $qqT$ operator is $\ep_i$ [see Eqs. (\ref{comp-q}),
(\ref{comp-que}) for definition of these factors], and the up quark Yukawa
matrix (involved in the coupling $qY_Uu^ch_u$) in a family space has the form
\beq
\begin{array}{ccc}
 & {\begin{array}{ccc}
\hs{-0.2cm}u^c_1 \hspace{0.5cm} & \hspace{0.3cm}u^c_2&\hspace{0.3cm}
u^c_3
\end{array}}\\ \vspace{1mm}
Y_U=
\begin{array}{c}
 q_1\vs{0.1cm}\\ q_2\vs{0.1cm} \\ q_3
 \end{array}\!\!\!\!\!\!\!\hs{-0.4cm} &{\left(\begin{array}{ccc}

\hs{0.3mm} a_1&\hs{0.1cm} a_{12} &\hs{0.1cm} a_{13}
\\
\vs{-0.4cm}
\\
 \hs{0.3mm}a_{21}& \hs{0.1cm} a_{2} & \hs{0.1cm} a_{23}
 \\
 \vs{-0.4cm}
 \\
 \hs{0.3mm}a_{31}&\hs{0.1cm}a_{32}&\hs{0.1cm}a_{3}

\end{array}\right)},~
\end{array}
\la{Yu}
\eeq
then the coupling $Y_{qq}$ (involved in $qY_{qq}qT$) will be
\beq
\begin{array}{ccc}
 & {\begin{array}{ccc}
\hs{-0.2cm}q_1 \hspace{0.6cm} & \hspace{0.75cm}q_2&\hspace{0.75cm} q_3
\end{array}}\\ \vspace{1mm}
Y_{qq}\simeq
\begin{array}{c}
 q_1\vs{0.1cm}\\ q_2\vs{0.1cm} \\ q_3
 \end{array}\!\!\!\!\!\!\!\hs{-0.4cm} &{\left(\begin{array}{ccc}

\hs{0.3mm} \ep_1a_1&\hs{0.15cm} \ep_{12}\bar a_{12} &\hs{0.1cm}
\ep_{13}\bar a_{13}
\\
\vs{-0.35cm}
\\
 \hs{0.3mm}\ep_{12}\bar a_{12}& \hs{0.15cm} \ep_2a_{2} & \hs{0.1cm}
\ep_{23}\bar a_{23}
 \\
 \vs{-0.35cm}
 \\
 \hs{0.3mm}\ep_{13}\bar a_{13}\hs{0.1cm}&\hs{0.15cm}
\ep_{23}\bar a_{23}&\hs{0.1cm}\ep_3a_{3}

\end{array}\right)},~
\end{array}
\la{Yqq}
\eeq
$$
{\rm with}~~~~\ep_{12}\bar a_{12}=\fr{1}{2}\l a_{12}\ep_2+a_{21}\ep_1\r ~,
$$
$$
\ep_{13}\bar a_{13}=\fr{1}{2}\l a_{13}\ep_3+a_{31}\ep_1\r ~,~~
\ep_{23}\bar a_{23}=\fr{1}{2}\l a_{23}\ep_3+a_{32}\ep_2\r ~.
$$
Note that since $q$ and $e^c$ states come from different $SU(5)$ states,
we can also avoid the asymptotic relation $\hat{M}_D=\hat{M}_E$ common for minimal
$SU(5)$ GUT. This will be discussed in more detail later on.

\subsection{Suppressing $ql\bar T$ and $u^cd^c\bar T$ operators}\label{sec:qlTsuppr}

Now we will present the mechanism for suppressing $ql\bar T$ couplings.
Recall that in $SU(5)$ this type of terms  originate from the couplings responsible for
generation of down quark and charged lepton masses [see Eq. (\ref{hTcoupl})].
The suppression of $ql\bar T$  can occur if the light $l$ and
$d^c$ are coming from different $SU(5)$ states. To realize such a splitting
in a natural way we introduce additional vector like $SU(5)$ matter
$\bar 5'+5'$, $\Psi(50)+\bar{\Psi }(\bar{50})$ and the following interaction terms
\beq
M_5\bar 5\cdot 5'+\rho \fr{\Si^2}{M_*}\bar 5'\Psi+
\bar{\rho }\fr{\Si^2}{M_*}5'\bar{\Psi }+M_{\Psi }\bar{\Psi }\Psi ~,
\la{5-Psi-coupl}
\eeq
($\rho $, $\bar{\rho }$ are dimensionless couplings).
The $50$-plet does not contain the state with the quantum
number of the lepton doublet \cite{ft3}, however it 
includes the state with quantum numbers of $d^c$. Therefore, after
substituting appropriate VEVs in (\ref{5-Psi-coupl}), for the mass couplings of the
corresponding  fragments we will have
\beq
\begin{array}{ccc}
 & {\begin{array}{cc}
\hs{-0.5cm}\bar d^c_{5'} \hspace{1cm} & \hspace{0.9cm}\bar d^c_{\Psi}
\end{array}}\\ \vspace{1mm}

\begin{array}{c}
 d^c_{\bar 5}\vs{0.1cm}\\ d^c_{\bar 5'}\vs{0.1cm} \\ d^c_{\bar{\Psi}}
 \end{array}\!\!\!\!\!\hs{-0.2cm} &{\left(\begin{array}{ccc}

\hs{0.6mm} M_5&\hs{0.3cm} 0
\\
\vs{-0.3cm}
\\
 \hs{0.6mm}0& \hs{0.3cm}\rho M_G\ep_G
 \\
 \vs{-0.3cm}
 \\
 \hs{0.6mm}\bar{\rho }M_G\ep_G\hs{0.1cm}&\hs{0.3cm}M_{\Psi }

\end{array}\right)},~~~~
\end{array}
\begin{array}{cc}
 & {\begin{array}{c}
\hs{-0.2cm}\bar l_{5'}
\end{array}}\\ \vspace{1mm}

\begin{array}{c}
l_{\bar 5'}\vs{0.1cm}\\ l_{\bar 5}
 \end{array}\!\!\!\!\!\hs{-0.02cm} &{\left(\hs{-0.15cm}\begin{array}{ccc}
0
\\
\vs{-0.3cm}
\\
M_5

\end{array}\hs{-0.15cm}\right)}~,
\end{array}
\label{compon5-50}
\end{equation}
where $\ep_G\equiv M_G/M_*$. As we see,  $l_{\bar 5}$ forms massive
state with $\bar l_{5'}$ and therefore the light lepton doublet emerges
from $\bar 5'$. However, the situation is different for $d^c$.
After integrating out $d^c_{\bar{\Psi}}$, $\bar d^c_{\Psi }$ states, the
$(2,1)$
element in the first matrix of (\ref{compon5-50}) receives the correction
$\tl{M}=M_G^2\ep_G^2/M_{\Psi}$. Assuming that $\tl{M}\gg M_5$, the light
$d^c$ state mostly remains in $\bar 5$, while light lepton doublet $l$ purely in $\bar 5'$. 
Therefore, we have
$$
\bar 5\supset d^c~,~~~\bar 5'\supset l~,
\ep''d^c~,
$$
\beq
\ep''=\fr{M_5}{\tl{M}}\ll 1~,~~~\tl{M}\sim \rho \bar{\rho }\fr{M_G^2}{M_{\Psi }}\ep_G^2~.
\la{dcl-weights}
\eeq
The masses of the decoupled states are
\beq
M(d^c_{\bar 5'}, \bar d^c_{5'})=\tl{M}~,~~~~M(l_{\bar 5}, \bar l_{5'})=M_5~,
\la{dec5}
\eeq
and all states from $\Psi $, $\bar{\Psi}$ have mass $M_{\Psi }$. From
(\ref{dcl-weights}) we see that the light lepton doublet and $SU(2)_L$ singlet
down quark are coming from different $SU(5)$ multiplets. This splitting will
be crucial for suppression of $ql\bar T$ coupling. To see this, we should
discuss the mass generation of the down quarks and charged leptons.
Thus, it is important to know where the light left handed quark doublet $q$ comes from. If the light $q$ state emerges from  $15$-plet and $e^c$ state from
$10$  [the mechanism ensuring suppression of $qqT$ coupling for
$1^{\rm st}$ or/and $2^{\rm nd}$ family; see Eq. (\ref{up-qqT})], then the
operators responsible for down quark and charged lepton masses are
$15\cdot \bar 5\bar H$ and $10\cdot \bar 5'\bar H$ respectively. Namely,
taking into account (\ref{comp-q}), (\ref{comp-ue}) and (\ref{dcl-weights})
we have
\beq
Y_D15\cdot \bar 5\bar H\to Y_Dqd^ch_d~,
\la{down-ql-1}
\eeq

\beq
Y_E10\cdot \bar 5'\bar H\to Y_E\l e^clh_d+\ep ql\bar T+\ep''u^cd^c\bar T\r ~.
\la{down-ql}
\eeq
As we see $ql\bar T$ term  emerges from coupling responsible for the charged
lepton mass and is suppressed by factor $\ep $. At the same time, the
$u^cd^c\bar T$ coupling is also suppressed. Since the $e^cu^cT$ coupling can be 
absent (see the discussion in the previous subsection) the corresponding right handed $d=5$ operator
$u^cu^cd^ce^c$ would not emerge at all. As far as the left handed operator is concerned, 
taking into account
(\ref{up-qqT}) and (\ref{down-ql}), it will have the form
\beq
\ep^2\fr{Y_UY_E}{M_T}qqql~.
\la{suppr-qqql}
\eeq
Note that together with suppression of $ql\bar T$, also the relation
$\hat{M}_D=\hat{M}_E$ is avoided. The reason is simple: the Yukawa matrices
$Y_D$ and $Y_E$ arise from completely independent interaction terms of (\ref{down-ql-1})
and (\ref {down-ql}) respectively.

Now let us show how the suppression of $ql\bar T$ coupling works for the case
corresponding to Eq. (\ref{comp-que}) (suppression of $qqT$ operator involving
third family). In this case the terms $10'\cdot \bar 5\bar H$ and
$10\cdot \bar 5'\bar H$ are responsible for down type quark and charged lepton
masses respectively. In particular, taking into account
(\ref{comp-que}), (\ref{dcl-weights}) we have
$$
Y_D10'\cdot \bar 5\bar H\to Y_Dqd^ch_d~,
$$
\beq
Y_E10\cdot \bar 5'\bar H\to Y_E\l e^clh_d+\ep'ql\bar T+\ep''u^cd^c\bar T\r ~.
\la{down-ql-1}
\eeq
Therefore, the corresponding $d=5$ operator emerging from (\ref{up-qqT-1}) and
(\ref{down-ql-1})
 \beq
(\ep')^2\fr{Y_UY_E}{M_T}qqql~,
\la{suppr-qqql-1}
\eeq
is suppressed by factor $(\ep')^2$, while $u^cu^cd^ce^c$-type operator is
still absent.

As we see, in both cases [corresponding to (\ref{suppr-qqql}) and
(\ref{suppr-qqql-1})] the $ql\bar T$ term emerges from the Yukawa couplings
responsible for charged lepton masses.
Thus, if $Y_E$ in a family space has the structure

\beq
\begin{array}{ccc}
 & {\begin{array}{ccc}
\hs{-0.2cm}l_1 \hspace{0.5cm} & \hspace{0.3cm}l_2&\hspace{0.3cm}
l_3
\end{array}}\\ \vspace{1mm}
Y_E=
\begin{array}{c}
 e^c_1\vs{0.1cm}\\ e^c_2\vs{0.1cm} \\ e^c_3
 \end{array}\!\!\!\!\!\!\!\hs{-0.4cm} &{\left(\begin{array}{ccc}

\hs{0.3mm} b_1&\hs{0.1cm} b_{12} &\hs{0.1cm} b_{13}
\\
\vs{-0.4cm}
\\
 \hs{0.3mm}b_{21}& \hs{0.1cm} b_{2} & \hs{0.1cm} b_{23}
 \\
 \vs{-0.4cm}
 \\
 \hs{0.3mm}b_{31}&\hs{0.1cm}b_{32}&\hs{0.1cm}b_{3}

\end{array}\right)},~
\end{array}
\la{YD}
\eeq
then the matrix $Y_{ql}$ (involved in $qY_{ql}l\bar T$ coupling) will be
\beq
\begin{array}{ccc}
 & {\begin{array}{ccc}
\hs{-0.2cm}l_1 \hspace{0.6cm} & \hspace{0.75cm}l_2&\hspace{0.75cm} l_3
\end{array}}\\ \vspace{1mm}
Y_{ql}\simeq
\begin{array}{c}
 q_1\vs{0.1cm}\\ q_2\vs{0.1cm} \\ q_3
 \end{array}\!\!\!\!\!\!\!\hs{-0.4cm} &{\left(\begin{array}{ccc}

\hs{0.3mm} \ep_1b_1&\hs{0.15cm} \ep_1b_{12} &\hs{0.1cm}
\ep_1b_{13}
\\
\vs{-0.35cm}
\\
 \hs{0.3mm}\ep_2b_{21}& \hs{0.15cm} \ep_2b_{2} & \hs{0.1cm}
\ep_2b_{23}
 \\
 \vs{-0.35cm}
 \\
 \hs{0.3mm}\ep_3b_{31}\hs{0.1cm}&\hs{0.15cm}
\ep_3b_{32}&\hs{0.1cm}\ep_3b_3

\end{array}\right)}~.
\end{array}
\la{Yql}
\eeq
Here,  the factors $\ep_i$  are the same as appeared in (\ref{Yqq}).

As we see, the split multiplet mechanisms we have discussed give good chance for the suppression of
nucleon decay. Of course, one should make sure that all couplings which may lead to fast proton decay are
absent. For example, the term $u^ce^cT$ can originate from the operator $10\cdot 10H$. Therefore, some care should be
taken to suppress such a coupling. Also, the Planck (cut off) scale suppressed $d=5$ baryon number violating operators must be
adequately suppressed. In a concrete model, presented in sect. \ref{sec:realSU5}, we will show that all this can be achieved and
justified by symmetry arguments.

\section{Naturally Suppressed $d=6$ Proton Decay}\label{sec:d6suppr}

In SUSY $SU(5)$ the exchange of super-heavy $V_X, V_Y$ gauge superfields
induce dimension six baryon number violating operators.
The corresponding $D$-terms are
$(qqu^{c\da }e^{c\da })_D$ and $(qlu^{c\da }d^{c\da })_D$.
 Dimension six operators also emerge in non SUSY GUTs and may be more problematic if the GUT scale
is lower than one in SUSY GUT ($\sim 10^{16}$~GeV with MSSM spectrum below $M_G$ scale).

Thanks to the mechanism discussed in the previous section, these kind of
operators can be also suppressed. Crucial role is played by splitting of
appropriate matter. We will discuss the $d=6$ operator suppression on example of SUSY $SU(5)$.
 Let us start consideration with $\bar 5$-plet
superfields which include states with the quantum numbers of $d^c$ and $l$.
The $D$-terms including $\bar 5$-plets are
\beq
\l \bar 5^{\da }e^{gV}\bar 5+\bar 5^{'\da }e^{gV}\bar 5'\r_D ~,
\la{D-5-pl}
\eeq
where $V$ and $g$ are $SU(5)$ gauge superfield and the gauge coupling at
scale $M_G$ respectively. According to (\ref{dcl-weights}), the $\bar 5$ states do
not include
light lepton doublets $l$  at all and therefore the first term in (\ref{D-5-pl}) is
irrelevant for us. However, from the second term of (\ref{D-5-pl}) we get
\beq
\l \bar 5^{'\da }e^{-gV}\bar 5'\r_D\to \ep''g\l l^{\da }V_Xd^c+
d^{c\da }V_Yl\r_{D}~.
\la{5-V-coup}
\eeq
As we see, the couplings of the heavy $V_{X,Y}$ gauge superfields
with the matter are suppressed by factor $\ep''$.

The kinetic $D$-term of $15$-plet is irrelevant for the baryon number violation because
from light states $15$-plet includes only $q$. For the case
corresponding to Eqs. (\ref{comp-q}), (\ref{comp-ue}) only $10$-plet's
$D$-term is relevant:
$$
(10^{\da }e^{gV}10)_D\to
$$
\beq
\ep g\l V_X(q^{\da }e^c+qu^{c\da })+V_Y(qe^{c\da }+q^{\da }u^c)\r_D ~,
\la{D-10-pl}
\eeq
producing couplings with the suppression factor $\ep $.

Upon integration of the $V_X, V_Y$ states with mass$\simeq M_G$ from
(\ref{5-V-coup}) and (\ref{D-10-pl}) we get the following baryon number
violating $d=6$ operators
\beq
\fr{g^2}{M_G^2}\left [ \ep^2 qqu^{c\da }e^{c\da }+
\ep \ep''qlu^{c\da }d^{c\da }+{\rm h.c.}\right ]_D~.
\la{d6ops}
\eeq
As we see, two $d=6$ operators in (\ref{d6ops}) are naturally suppressed by
factors $\ep^2$ and $\ep \ep''$ respectively. Note  that if we are dealing
with case corresponding to Eq. (\ref{comp-que}), then the factor $\ep $ in
(\ref{d6ops}) must be replaced by $\ep'$. Once more, this mechanism for the suppression of
$d=6$ nucleon decay also can be applied within non SUSY $SU(5)$.

\section{Example of Realistic SUSY $SU(5)$}\label{sec:realSU5}

The possibilities for suppressing the proton decay in SUSY $SU(5)$ GUT
discussed above can be successfully applied for the realistic model building.
By proper selection of the appropriate mass scales we can get suppression
[$\ep $, $\ep'$ in Eqs. (\ref{comp-q}), (\ref{comp-que})] as strong as we
wish.
This requires the scales $M_{15}$ and $M_{5}$ to be below $M_G$.
However, this introduces an additional states below the GUT scale, and the
running of gauge couplings will be altered. In order to maintain successful
gauge coupling unification, an  additional constraint on these scales
should be imposed. Suppression of
$qqT$ coupling brings the states $(S+\bar S)_{15}$ and $(\De +\bar{\De })_{15}$
below $M_G$. With their  masses $M_S=M_{\De }=M_{15}$, one can
see that the states $S, \bar S$ contribute stronger to the running of $\al_3$
in comparison of $\De , \bar{\De }$'s contribution into the $\al_2$ running.
To compensate this dis-balance an additional $SU(2)_L$ states are required.
This occurs naturally
if the mechanism for $ql\bar T$ suppression is invoked. In this case
below $M_G$ we also have an additional $SU(2)_L$ doublets
(see Eq. (\ref{dec5})). This offers possibility for successful gauge coupling
unification.

Now, we present an example of SUSY $SU(5)$ realizing ideas discussed
above. Considering three families of quarks and leptons, the appropriate
couplings (such as of Eqs. (\ref{10-15}), (\ref{up-qqT}),
(\ref{5-Psi-coupl}), (\ref{down-ql}))
should be promoted to the matrices in a family space.

Thus, we introduce three pairs of $15$-plets: $(15+\bar{15})_i$
($i=1, 2, 3$) and the pair $10'+\bar{10}'$ (needed for renormalizable top
Yukawa couping), and also $(\bar 5'+5')_i$,
$(\Psi +\bar{\Psi})_i$.

In addition, we introduce ${\cal U}(1)\tm {\cal Z}_{3N}$ symmetry, where as will turn out ${\cal U}(1)$
is an anomalous and ${\cal Z}_{3N}$ is discrete symmetry. Importance of these symmetries will become obvious
soon. The anomalous $U(1)$ factors can appear in effective field theories from strings and
cancellation of its anomalies occurs through the Green-Schwarz mechanism \cite{Green:1984sg}. Due to the
anomaly, the Fayet-Iliopoulos term $-\xi \int d^4\te V_A$ is always generated \cite{Witten:1981nf} and the corresponding
$D_A$-term  has the form \cite{Dine:1987xk}
\beq
\fr{g_A^2}{8}D_A^2=\fr{g_A^2}{8}\l -\xi +\sum Q_i|\phi_i|^2\r^2~,~~~\xi =\fr{g_A^2M_P^2}{192\pi^2}{\rm Tr}Q~,
\la{FI-D-A}
\eeq
where $Q_i$ is the ${\cal U}(1)$ charge of superfield  $\phi_i$. The transformations under ${\cal U}(1)$ and ${\cal Z}_{3N}$
are respectively
$$
{\cal U}(1):~~~\phi_i\to e^{{\rm i}Q_i}\phi_i~,
$$
\beq
{\cal Z}_{3N}:~~~\phi_i\to e^{{\rm i}q_i\om }\phi_i~,~~{\rm with}~~\om =\fr{2\pi }{3N}~.
\la{trans-u1-ZN}
\eeq
The anomalous ${\cal U}(1)$ can be very useful for building models with realistic phenomenology \cite{Dvali:1996sr} ,
and we also take advantage of it here for avoiding unwanted couplings. The symmetry ${\cal Z}_{3N}$ also will play a crucial role.
We introduce two $SU(5)$ singlet superfields $X$ and $Z$
which will be used for ${\cal U}(1)\tm {\cal Z}_{3N}$ breaking. The $Q_i$ and $q_i$ charges of scalar superfields
are given in Table \ref{t:1}.
%
%
%
%
\begin{table} \caption{${\cal U}(1)\tm {\cal Z}_{3N}$ charges $Q, q$ of the scalar superfields.}

\label{t:1} $$\begin{array}{|c||c|c|c|c|c|}

\hline
\vs{-0.2cm}
 &  &  &  & &    \\

\vs{-0.3cm}

& ~X~& ~Z ~& ~\Si (24) ~  & ~H(5)~ & ~\bar H(\bar 5) ~   \\

&  &  &  & &    \\

\hline
\vs{-0.15cm}
&  &  &  & &    \\

\vs{-0.15cm}

~Q~& 1 & -1 &0  &-1/3&-2/3    \\

&  &  &  & &  \\

\hline
\vs{-0.15cm}
&  &  &  & &    \\

\vs{-0.15cm}

~q~& 0 &3  &0  &-1 &1    \\

&  &  &  & &  \\

\hline

\end{array}$$

\end{table}
%
%
Let us first discuss the VEV generation for scalar components of $X$ and $Z$ superfields. The lowest superpotential
coupling for these superfields , allowed by ${\cal U}(1)\tm {\cal Z}_{3N}$
symmetry, is
\beq
W(X, Z)=\si M_*^3\l \fr{XZ}{M_*^2}\r^N~,
\la{S-sup}
\eeq
where $\si $ is dimensionless coupling.
With $\xi >0$, in unbroken SUSY limit the conditions 
$D_A=0$, $F_X=F_Z=0$ give
$\lan X\ran =\sq{\xi }$ and $\lan Z \ran =0$.
However, the non zero VEV for $Z$ can be generated after including the
soft SUSY breaking potential terms
\beq
V_{SB}=m_{3/2}^2\l |X|^2+|Z|^2\r -Am_{3/2}\l W+W^{\dagger }\r
\la{V-SB}
\eeq
Thus, we should minimize the whole potential
\beq
V=\fr{g_A^2}{8}D_A^2+|F_X|^2+|F_Z|^2+V_{SB}~,
\la{V-tot}
\eeq
where
\beq
D_A=-\xi +|X|^2-|Z|^2~,~~F_X=\fr{\pl W}{\pl X}~,~~~F_Z=\fr{\pl W}{\pl Z}~.
\la{DA-FX-FZ}
\eeq
Considering soft breaking contribution as a perturbation to the potential's leading part, it is natural that 
by proper selection of $N$ we will get $\lan Z\ran \ll \lan X\ran $. Therefore we parameterize the vacuum as
\beq
\lan |X|^2\ran =\xi (1-\ka^2)+\al m_{3/2}^2~,~~~\lan |Y|^2\ran =\ka^2\xi ~,
\la{vac-par}
\eeq
where
\beq
\al \sim 1~,~~~\ka \ll 1~,
\la{al-ka-cond}
\eeq
should be found from the minimization. Note that with this parameterization $D_A$-term is shifted by the SUSY scale$\sim m_{3/2}^2$.
Minimizing the potential (\ref{V-tot}) with real $A$ and conditions in (\ref{al-ka-cond}) we find analytically
$$
\al=-\fr{4}{g_A^2}+{\cal O}(\ka^2)~,~~~\ka =\ov{\al } \l \fr{m_{3/2}}{M_*}\r^{\fr{1}{N-2)}}\l \fr{\sq{\xi }}{M_*}\r^{-2\fr{N-1}{N-2}}~,
$$
\beq
{\rm with }~~\ov{\al }=\l \fr{A\pm \sq{A^2-8(N-1)}}{2\si N(N-1)}\r^{\fr{1}{N-2}}~.
\la{sol-al-ka}
\eeq
With  $m_{3/2}\sim 1$~TeV, $M_*\sim 10^{17}$~GeV (we will comment on this value of the cut off  scale below), $\sq{\xi }\sim 10^{16}$~GeV
and $N=6$ we  have $\ka \sim 0.1$ and therefore initial assumption (\ref{al-ka-cond}) is justified. Thus, finally we have
\beq
\fr{\lan X\ran }{M_*}\simeq \fr{\sq{\xi }}{M_*}=0.1~,~~~ \fr{\lan Z\ran }{M_*}\simeq \ka \fr{\sq{\xi }}{M_*} \sim 10^{-2}~.
\la{VEV-SX}
\eeq
Below we will use these values obtained by ${\cal U}(1)\tm {\cal Z}_{3N}$  
($N=6$) symmetry.

One may wonder whether with charge assignments given is Table \ref{t:1}, desirable GUT symmetry breaking
can occur or not. Also, the color triplets from $H, \bar H$ should be super-heavy, while doublets
should remain massless (doublet-triplet (DT) splitting).
Since the adjoint $\Si $ is not transformed under ${\cal U}(1)\tm {\cal Z}_{3N}$ symmetry, the renormalizable superpotential
includes couplings 
\beq
W(\Si )=M_{\Si }{\rm Tr}\Si^2+\lam_{\Si }{\rm Tr}\Si^3~,
\la{sup-Sigma}
\eeq 
and the non zero
VEV $\lan \Si \ran =V_{\Si }\cdot {\rm Diag}\l 2, 2, 2, -3, -3\r $ with $V_{\Si }=\fr{2M_{\Si }}{3\lam_{\Si }}$ is
obtained. This insures the breaking $SU(5)\to SU(3)_c\tm SU(2)_L\tm U(1)_Y$.
As far as the DT
splitting is concerned, without invoking some particular mechanism it can be achieved by fine tuning \cite{ftDT}
if couplings $(M_H+\lam_H\Si )\bar HH$ exist. However, these couplings 
are forbidden by ${\cal U}(1)\tm {\cal Z}_{3N}$ symmetry.
Instead, the operators $\lam' X\bar HH+X\Si \bar HH/M'$ are allowed. 
They lead to the DT splitting with $M'\sim \lan X\ran $ and
$\lam'\sim V_{\Si }/\lan X\ran $. The operator $X\Si \bar HH/M'$ can be generated by decoupling of additional states with mass $M'$.
For example, introducing $H'(5), \bar H'(\bar 5)$ states with $(Q, q)$ charges $(-1/3, -1)$ and $(1/3,1)$ respectively, the relevant couplings
are
\beq
\lam'X\bar HH+\si_1\Si \bar H'H+\si_2X\bar HH'+M'\bar H'H'~.
\la{H1-coup}
\eeq
After integrating out the heavy $H', \bar H'$ states, we remain with effective superpotential couplings
\beq
\lam'X\bar HH-\si_1\si_2\fr{X}{M'}\Si \bar HH~.
\la{WH-eff}
\eeq
With a selection $\lam'=-3\si_1\si_2V_{\Si }/M'$ the MSSM Higgs doublets remain massless, while the color triplets occur
mass $M_T=5\lam'\lan X\ran /3$($\sim M_G$ with $\lam'\sim V_{\Si }/\lan X\ran $). Therefore, we see that within SUSY $SU(5)$ GUT, augmented with ${\cal U}(1)\tm {\cal Z}_{3N}$ symmetry, it is possible to
build self consistent scalar sector.

Now we are ready to discuss the fermion sector. The ${\cal U}(1)\tm {\cal Z}_{3N}$ charge assignments for matter states are
displayed in Table \ref{t:2}. Note that with this prescription all matter parity violating operators are forbidden. Therefore,
thanks to the ${\cal U}(1)\tm {\cal Z}_{3N}$ symmetry the $R$-parity is automatic. The reason for this is the fact that by VEVs 
$\lan X\ran , \lan Z\ran $ the ${\cal U}(1)\tm {\cal Z}_{3N}$ is not completely broken. Namely, the subgroup ${\cal Z}_3^A\tm {\cal Z}_3$
remains unbroken. The transformations under ${\cal Z}_3^A$ and ${\cal Z}_3$
 are respectively
$$
{\cal Z}_3^A:~~~\phi_i\to e^{{\rm i}2\pi Q_i}\phi_i~,
$$
\beq
{\cal Z}_3:~~~\phi_i\to e^{{\rm i}q_i\ov{\om }}\phi_i~,~~{\rm with}~~\ov{\om }=\fr{2\pi }{3}~.
\la{trans-u1-ZN}
\eeq
The superfields $X, Z$ are neutral under ${\cal Z}_3^A$ and ${\cal Z}_3$.

%
%
%
\begin{table} \caption{${\cal U}(1)\tm {\cal Z}_{3N}$ charges $Q, q$ of matter superfields.}

\label{t:2} $$\begin{array}{|c||c|c|c|c|c|c|c|}

\hline
\vs{-0.2cm}
 &  &  &  & &  & &  \\

\vs{-0.3cm}

&~ 10_i~ &~ \bar {15}_i~& ~\bar 5_i~, 5'_i~& ~\bar 5'_i ~, \bar{\Psi }_i~&~ 15_i ~, 10' ~&~ \bar{10}'~&~\Psi_i  ~\\

&  &  &  & &  & &  \\

\hline
\vs{-0.15cm}
&  &  &  & &   & & \\

\vs{-0.15cm}

~Q~& -1/3& 1/3 &0  &1 &2/3   &-2/3 &-1 \\

&  &  &  & &  & &\\

\hline
\vs{-0.15cm}
&  &  &  & &  & &  \\

\vs{-0.15cm}

~q~& 2 &-2  &0  &-3 &-1   &1 &3 \\

&  &  &  & & & & \\

\hline

\end{array}$$

\end{table}
%
%

 In understanding of observed hierarchies between charged fermion masses and mixings crucial role plays the flavor
structure of the Yukawa sector. Same is true in connection of the color higgsino mediated and Planck scale
suppressed $d=5$ operator induced nucleon decay. Their structures determine the signature of the nucleon decay .
Definite structures as well as predictions can be obtained by flavor symmetries. Indeed, symmetry principle is 
very powerful for a predictive power. We will not introduce here generation symmetries, and instead consider one 
particular example demonstrating realization of the split multiplet mechanism.

Thus we promote the  couplings of (\ref{10-15}), (\ref{10'-10-15})  in the flavor space as
\beq
\lam_i10_i\Si \bar{15}_i+\bar{\lam }15_3\Si \bar{10}'+\lam_i^ZZ\bar{15}_i15_i+
M_{10}10'\bar{10}'~,
\la{fl-15-15}
\eeq
where for simplicity we have assumed that the matrices $\lam $, $\lam^Z$ are diagonal and
only $15_3$ couples with $\bar{10}'$. Moreover, we take
$$
\lam_i\sim \bar{\lam }\sim 1~,~~~M_{10}\sim M_{G}~,
$$
\beq
M_{15_1},~M_{15_2}=M_{15_3}\equiv M_{15}\ll M_G~,
\la{real-su5-cond}
\eeq
where $M_{15_i}=\lam^Z_i\lan Z\ran $.
Thus, with
$$
\ep_1=\fr{M_{15_1}}{\lam_1M_G}~,~~~~~\ep_2=\fr{M_{15_2}}{\lam_2M_G}~,
$$
\beq
\ep_3=\fr{M_{10}M_{15_3}}{\lam_3\bar{\lam }M_G^2}~,~~~~(\ep_2\sim \ep_3\equiv \ep)~,
\la{real-su5-eps}
\eeq
and carrying out analysis analogous done in sect. \ref{sec:qqTsuppr}, we will have
$$
q_{1, 2}\subset 15_{1, 2}~,~~~q_3\subset 10'~,~~~~(u^c, e^c)_{1, 2, 3}\subset 10_{1, 2, 3}~,
$$
\beq
10_1\supset \ep_1q_1~,~~~~10_2\supset \ep q_2~,~~~~10_3\supset \ep q_3~.
\la{realsu5-q-weights}
\eeq
The couplings $\fr{Z}{M_*}10'\Si \bar{15}_i$ because of the suppression factor$\sim\lan Z\ran/M_*\sim 10^{-2}$ do not change these relations.
They cause  $15_i\stackrel{\supset }{_\sim }10^{-3}q_3$, $10_i\stackrel{\supset }{_\sim }10^{-2}q_3$ which will not have any impact on our studies.
{}The mass spectrum of the decoupled states is
$$
M(q_{10_i}, \bar q_{\bar{15}_i})\sim M(q_{15_3}, \bar q_{\bar{10}'})\sim
M(u^c_{10'}, \bar u^c_{\bar{10}'})\simeq
$$
$$
M(e^c_{10'}, \bar e^c_{\bar{10}'})
\simeq M_G~,~~~
M_{S_1}=M_{\De_1}=M_{15_1}\simeq \ep_1M_G~,
$$
\beq
M_{S_{2,3}}=M_{\De_{2,3}}=M_{15}\simeq \ep M_G~.
\la{real-su5-spectrum1}
\eeq
Moreover, the couplings in (\ref{5-Psi-coupl}) will be replaced by ${\cal U}(1)\tm {\cal Z}_{3N}$ invariant terms
\beq
\begin{array}{ccc}
 & {\begin{array}{cc}
\hs{-0.1cm}{5_i}' \hspace{1.3cm} & \hspace{1.3cm}\Psi_i
\end{array}}\\ \vspace{1mm}

\begin{array}{c}
 \bar 5_i\vs{0.1cm}\\ \bar {5'}_i\vs{0.1cm} \\ \bar{\Psi }_i
 \end{array}\!\!\!\!\!\hs{-0.2cm} &{\left(\begin{array}{ccc}

\hs{0.6mm} M_5&\hs{0.3cm} 0
\\
\vs{-0.3cm}
\\
 \hs{0.6mm}{M_5}'Z/M_*& \hs{0.3cm}\rho \Si^2/M_*
 \\
 \vs{-0.3cm}
 \\
 \hs{0.6mm}\bar{\rho }\Si^2Z/M_*^2\hs{0.1cm}&\hs{0.3cm}M_{\Psi }

\end{array}\right)}~.
\end{array}
\label{real-5-Psi-coupl}
\end{equation}
Here we still assumed that the appropriate entries
are diagonal and universal. We assume that $M_5\sim {M_5}'$ (the smallness 
of both these scales, with respect to
$M_G$ or $M_*$, may have same origin. However, this is not explained here).
Therefore, carrying out similar analysis presented in previous section, with
\beq
\ep''\equiv \fr{M_5}{\tl{M}}\ll 1~,~~~\tl{M}\sim \rho \bar{\rho }\fr{M_G^2\lan Z\ran }{M_{\Psi }M_*}\ep_G^2~,
\la{real-cond-M5-tlM}
\eeq
we have
\beq
\bar 5_i\supset d^c_i~,10^{-2}l_i~~~
\bar {5'}_i\supset l_i~, \ep''d^c_i~.
\la{real-dcl-weights}
\eeq
The decoupled states will have the masses
$$
M(l_{\bar 5_i}, l_{{5'}_i})=M_5~,
$$
\beq
M(d^c_{\bar{5_i}'}, \bar d^c_{{5_i}'})=\tl{M}~,~~~~M_{\Psi_i}=M_{\Psi }~.
\la{real-su5-spectrum2}
\eeq
Note that in $(1,2)$ entry of (\ref{real-5-Psi-coupl}) the operator $\Si^2(XZ)^5/M_*^{11}$ is allowed.  However, it would
induce strongly suppressed correction$\sim 10^{-17}M_G $ and is not relevant.
Now we can discuss the gauge coupling unification.
The latter suggests the particular selection for appropriate mass scales.

\subsection{Gauge coupling unification}\label{sec:unif-real}

We assume that the masses of the matter $50_i$-plets are close to the cut off
scale $M_{\Psi }\simeq M_*$ - much higher than the GUT scale. Thus, they do not affect the gauge 
coupling running.  Moreover, with $M_{\Si }\sim M_G$ and $\lam_{\Si }\sim 1$ in (\ref{sup-Sigma})
for colored octet and $SU(2)_L$ triplet (from adjoint $\Si $) masses we get $m_8=m_3\sim M_G$ and
with $M_G\ll M_*$ higher order operators will not affect this relation. Also, with color triplets' mass
(from $H, \bar H$) near$\sim M_G$, these states will not  contribute to the gauge coupling running and
at the leading order will not play role in determination of the GUT scale (unlike the proposals of \cite{propos}).
However, for the masses of the three vector like pairs we have
$M(d^c_{\bar 5'}, \bar d^c_{5'})=\tl{M}$
(see Eq. (\ref{real-su5-spectrum2})). Apart these states, below $M_G$ we have
$3\tm (l_{\bar 5}+\bar l_{5'})$ and $3\tm (S+\bar S+\De +\bar{\De })_{15}$
states with masses given in (\ref{real-su5-spectrum1}) and (\ref{real-su5-spectrum2}).
Thus, for the strong gauge coupling constant at $M_Z$ scale  in 1-loop
approximation we get:
\beq
\al_3^{-1}=\l \al_3^0\r^{-1}-\fr{3}{2\pi }\ln \l \ep^2\ep_1\r
-\fr{27}{14\pi }\ln \fr{\tl{M}}{M_5}~,
\la{al3-Model}
\eeq
where $\al_3^0$ is the value of the strong coupling constant within MSSM
and is $\al_3^0(M_Z)\simeq 0.126$ \cite{Langacker:1995fk}.
The additional terms in (\ref{al3-Model}) allow to obtain the value
compatible with experiments $\al_3^{\rm exp}(M_Z)\simeq 0.1176$ \cite{Yao:2006px}.
This can be achieved with $\fr{M_5}{\tl{M}}\approx e^{14/15}\l \ep^2\ep_1 \r^{7/9}$.
In order to have more accurate estimate we have performed
calculations in two loop approximation. The picture of gauge coupling
unification is given in Fig. 1.
%

%
\begin{figure}
\begin{center}
\leavevmode
\leavevmode
\vspace{2.5cm}
\includegraphics{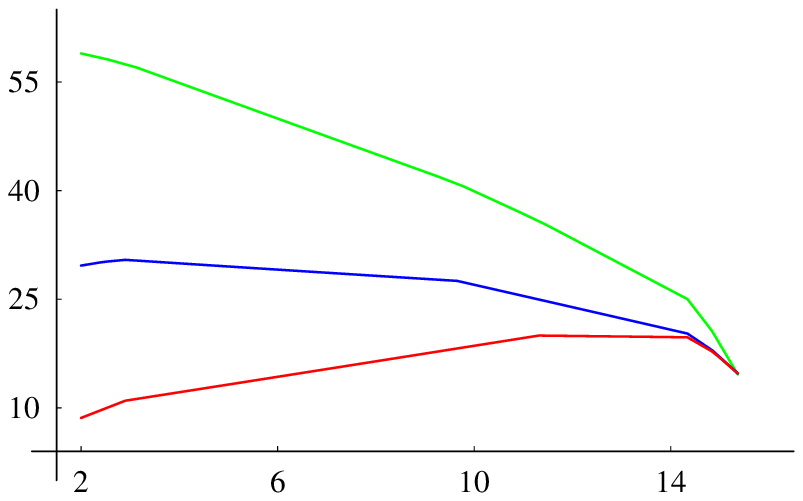}
\end{center}
\vs{3cm}
\caption{Gauge coupling unification. $\al_3(M_Z)\simeq 0.1176$,
$M_G\simeq 2.2\cdot 10^{15}$~GeV.
}
\rput(2.8,1.7){$\log_{10}\fr{\mu }{\rm GeV}$}
\rput(-3.7,4.5){$\al^{-1}$}
\end{figure}

For simplicity we have taken all squark, slepton, higgsino and gaugino masses
all near the TeV scale.
In particular
$$
m_{\tl{q}}=m_{\tl{l}}=M_{\tl{h}}=M_{\tl{g}}=
m_{susy}=10^{2.9}~{\rm GeV}~, 
$$
\beq
M_{\tl{W}}=M_{\tl{g}}\left. \fr{\al_2}{\al_3}\right |_{\mu =m_{susy}}\simeq 287~{\rm GeV}~. 
\la{soft-spectrum}
\eeq
Also, we have taken
$$
\ep_1=1/3~,~~~~~\ep=0.1~,~~~~~\fr{M_5}{\tl{M}}\simeq 2.2\cdot 10^{-2}~,
$$
\beq
{\rm with}~~~\tl{M}\simeq 2.1 \cdot 10^{11}~{\rm GeV}~.
\la{ratios-Model}
\eeq
 (For stronger suppression of nucleon decay smaller values of $\ep_1, \ep $ are required. However,
this would make new states lighter and the constraint from the coupling unification does not give much flexibility.)
All this and input values $\al_1^{-1}(M_Z)=59.0$, $\al_2^{-1}(M_Z)=29.6$ provides the
successful unification with
$$
\al_3(M_Z)=0.1176~,~~~~~M_G\simeq 2.2\cdot 10^{15}~{\rm GeV}~,
$$
\beq
\al_G^{-1}(M_G)\simeq 14.91~.
\la{al3-MG-alG-Model}
\eeq
As we see, due to the new states the unification scale $M_G$ is reduced, while the
unified gauge coupling $\al_G$ is enhanced:
\beq
\fr{M_G}{M_G^0}\simeq \fr{1}{9.1}~,~~~~~~~\fr{\al_G}{\al_G^0}\simeq 1.6~,
\la{MG-alG-ratios}
\eeq
[superscript $'0'$ indicates the values obtained within minimal SUSY $SU(5)$].
These values will be useful for estimating of the proton decay rates in
this model.

One can see that near $10^{17}$~GeV scale the unified gauge coupling
becomes strong  $\fr{\al_G}{4\pi }\simeq 0.26$. Thus, for cut off scale one should take $M_*\simeq 10^{17}$~GeV.
With this, the perturbative regime is kept in a quite wide range above the GUT scale.
As we will see shortly, the values of $\ep_1$ and $\ep $ selected here provide an adequate
suppression of the proton decay.

Finally, we calculate  short range renormalization
factors which will be used in the next subsection.
The appropriate baryon number violating $d=5$ and $d=6$ operators, generated at GUT scale, should
be defined at scale $\mu =1$~GeV. Thus two ranges are relevant for renormalization. Due to
running from $M_G$ down to $M_Z$ (or SUSY scale) the appropriate factor $A^S$ is called short range
renormalization factor. From scale $M_Z$ down to $1$~GeV the appearing factor $A_L$ is the long range
factor which is mainly  due to QCD running.

Let us start with calculation of the short range factor corresponding to $d=5$ operators.
Note that in our model the $ql\bar T$ coupling is related to the charged lepton Yukawa
matrix. Therefore,
generalizing expression given in \cite{Ibanez:1984ni}, we will have
$$
A_{d=5}^S=A_{d=5, ue}^S=A(\lam_t)\prod_{i; a>b}\l \fr{\al_i(\mu_a)}{\al_i(\mu_b)} \r^{\fr{c_i}{b_i(\mu_{a-b})}}~,
$$
\beq
c_i=\l -\fr{17}{15}, 0, \fr{4}{3}\r ~,
\la{AS-d5}
\eeq
where $b_i(\mu_{a-b})$ denotes gauge coupling 1-loop $b$-factors in the mass interval $\mu_b-\mu_a$ and
$A(\lam_t)$ includes the renormalization effect due to the top Yukawa coupling. We have evaluated $A_{d=5}^S$
for our scenario (more details of the Yukawa sector is given in section \ref{sec:lifetime-real}) in 2-loop
approximation for $\lam_t(M_Z)\simeq 1$ and obtained
\beq
A_{d=5}^S\simeq 2.03~,
\la{model-AS-d5}
\eeq
(to be compared with the factor obtained in MSSU5 $(A_{d=5}^S)^0=(A_{d=5, ud}^S)^0\simeq 0.92$).

As far as the $d=6$ operators are concerned, as it will turn out, the first type operator
of Eq. (\ref{d6ops}) will be relevant. It's 1-loop short range renormalization factor
is given by
\beq
A_{d=6}^S=\hs{-1mm}\prod_{i; a>b}\hs{-2mm}\l \fr{\al_i(\mu_a)}{\al_i(\mu_b)} \r^{\fr{\bar c_i}{b_i(\mu_{a-b})}}~,~~
\bar c_i=\l \fr{23}{30}, \fr{3}{2}, \fr{4}{3}\r .
\la{AS-d6}
\eeq
In our model numerically we get $A_{d=6}^S\simeq 2.23$. Also long range renormalization factor
$A_L$ should be taken into account. The latter is $A_L\simeq 1.34$ \cite{Nihei:1994tx},
and finally for $d=6$ operator renormalization factor we have
\beq
A^{d=6}_R=A_LA_{d=6}^S\simeq 2.99~.
\la{tot-AR-d6}
\eeq

\subsection{Proton life time}\label{sec:lifetime-real}

The nucleon decay via $d=5$ operators crucially depends on Yukawa sector.
Therefore, first we briefly
discuss how desirable fermion pattern can be obtained. Since we have arranged 
the multiplet splitting, displayed
in Eqs.  (\ref{realsu5-q-weights}), (\ref{real-dcl-weights}), will not be difficult
to get realistic fermion masses. 
 Once more we stress that we consider one particular example with simple flavor structure.
Starting with up type quarks, we will
write appropriate couplings in such a way that the up quark mass matrix will be diagonal. Relevant terms
consistent with ${\cal U}(1)\tm {\cal Z}_{3N}$ symmetry are
\beq
\fr{\Si }{M_*}\l \ga_115_110_1+\ga_215_210_2\r H+\ga_310'10_3H~,
\la{realsu5-up-yuk}
\eeq
where $\ga_{1,2,3}$ are dimensionless constants. Taking into account (\ref{realsu5-q-weights})
we will have
\beq
Y_U={\rm Diag}\l \lam_u~, \lam_c~, \lam_t\r ~,~~\lam_{u, c}\sim \ga_{1,2}\fr{\lan \Si \ran}{M_*}~,~\lam_t=\ga_3 .
\la{model-YU}
\eeq
Thus, the CKM mixings ($V_{ij}$) should come from the down quark sector.
The relevant couplings for the latter are
\beq
\l 15_1,~15_2,~10'\r Y_D\hs{-0.3cm}
\begin{array}{cc}
\vspace{1mm}

\begin{array}{c}
\vs{0.1cm}\\
 \end{array}\!\!\!\!\!\hs{-0.02cm} &{\left(\hs{-0.15cm}
 \begin{array}{ccc}
\bar 5_1
\\
\bar 5_2
\\
\bar 5_3

\end{array}\hs{-0.15cm}\right)}\vs{-0.2cm}\bar H~,
\end{array}
\label{realsu5-down-yuk}
\end{equation}
where the non-diagonal Yukawa matrix $Y_D$ is responsible for CKM mixings.

Thanks to the mechanism discussed in sect. \ref{sec:d5suppr}, the charged lepton Yukawa
matrix elements are independent from $Y_D$. For simplicity we take diagonal
couplings
\beq
Y_i10_i\bar {5'}_i\bar H~,
\la{realsu5-chlep-yuk}
\eeq
which with  (\ref{realsu5-q-weights}), (\ref{real-dcl-weights}), give
\beq
Y_E={\rm Diag}\l \lam_e~,~\lam_{\mu }~,~\lam_{\tau }\r ~,~~~\lam_{e, \mu ,\tau }=Y_{1,2,3}~.
\la{model-YE}
\eeq
Note, that the ${\cal U}(1)\tm {\cal Z}_{3N}$ symmetry provides the suppression of the coupling $15\cdot \bar 5'\bar H$
by factor$\sim \lan Z\ran/M_*\sim 10^{-2}$ in comparison of (\ref{realsu5-down-yuk}), (\ref{realsu5-chlep-yuk}) operators, and therefore
can be ignored. As far as the $10\cdot \bar 5\bar H$ type couplings are concerned, they are strongly 
suppressed($\sim \lan \fr{X(XZ)^5}{M_*^{11}}\ran \sim 10^{-16}$).
{}From all this and Eqs. (\ref{Yqq}), (\ref{Yql}), $Y_{qq}$ and
$Y_{ql}$ matrices will be
$$
Y_{qq}={\rm Diag}\l \ep_1\lam_u~,~\ep \lam_c~,~\ep \lam_t\r ~,
$$
\beq
Y_{ql}={\rm Diag}\l \ep_1\lam_e~,~\ep \lam_{\mu }~,~\ep \lam_{\tau }\r ~.
\la{model-Yqq-Yql}
\eeq
These couplings induce $qqql$ type $d=5$ left handed operators.

Before estimating the proton life time, let us note that the couplings $10_i10_jH$ are forbidden
by ${\cal U}(1)\tm {\cal Z}_{3N}$ symmetry. Only the higher order operators $\fr{Z^5X^6}{M_*^{11}}10_i10_iH$
are allowed. For the VEVs given in (\ref{VEV-SX}), induced operator $u^ce^cT$ is suppressed
by factor $\sim 10^{-16}$. This makes color triplet induced $d=5$ right handed baryon number violating operators completely
irrelevant. As far as the cut of scale suppressed $d=5$ baryon number violating operators are concerned, one can easily
see that they  also involve high powers of $\lan Z\ran /M_*$ and  $\lan X\ran /M_*$.
Namely, the allowed couplings are $X(XZ)^410_i10_j10_k\bar 5,~(XZ)^510_i10_j10_k\bar 5'$, etc.
Thus the suppression by factors$\stackrel{<}{_\sim }10^{-13}$ is guaranteed. Therefore, we conclude that in our model {\it only} sources for the proton decay
 are the couplings given in (\ref{model-Yqq-Yql}) and $X, Y$ boson induced decay which we discuss afterwords.

The appropriate $d=5$ left handed
operator is converted to four fermion operators through the Wino dressings.
Those, relevant for nucleon decay, have  forms
\beq
-\frac{1}{M_T}{\cal F}
\alpha_{ijk} (u d^i)(d^j\nu^k)~,
\label{d6nu}
\eeq
 \beq
\frac{1}{M_T}{\cal F}
{\alpha'}_{ij} (u d^i)(ue^j)~,
\label{d6e}
\eeq
where, together with other family independent factors, ${\cal F}$ includes
the loop integral, and for simplicity we have assumed that the squarks and
sleptons of all families have universal mass. The flavor dependent couplings
$\al_{ijk}$ and ${\alpha_{ij}}'$ are given by \cite{Nath:1988tx},
\cite{Berezhiani:1998hg}
$$
\al_{ijk}=
(L_d^{\da }Y_{ql}L_e)_{jk}(V^TL_u^{\da }Y_{qq}L_d^{*}V^{\da })_{i1}+
$$
$$
(L_d^{\da }Y_{qq}L_u^{*}V)_{ji}
(V^*L_d^{\da }Y_{ql}L_e)_{1k}-
(L_u^{\da }Y_{qq}L_d^{*})_{1i}(V^TL_u^{\da }Y_{ql}L_e)_{jk}
$$
\beq
+(L_d^{\da }Y_{qq}L_u^{*}V)_{ij}(L_u^{\da }Y_{ql}L_e)_{1k}~,
\label{d6nu1}
\eeq
$$
{\al'}_{ij} =
-(L_u^{\da }Y_{qq}L_d^{*})_{1i}(V^*L_d^{\da }Y_{ql}L_e)_{1j}+
$$
$$
(L_u^{\da }Y_{qq}L_d^{*}V^{\da })_{11}
(L_d^{\da }Y_{ql}L_e)_{ij}
+(L_u^{\da }Y_{ql}L_e)_{1j}(V^TL_u^{\da }Y_{qq}L_d^{*}V^{\da })_{1i}
$$
\beq
+(L_u^{\da }Y_{qq}L_d^{*}V^{\da })_{11}
(L_e^{T}Y_{ql}^TL_u^{*}V)_{j1}~.
\label{d6e1}
\eeq
$L_{u, d, e}$ are unitary matrices transforming the left handed fermion states
in order to diagonalize corresponding mass matrices.

{}For the considered case here we have $L_u=L_e={\bf 1}$, $L_d=V^*$.
Therefore, the only non-diagonal matrix is the CKM matrix. 
In particular, using (\ref{model-Yqq-Yql}),we have
$\al_{ijk}=2\de_{1k}\lam_e\ep_1(V^TY_{qq}V)_{ij}$. These factors
are responsible for the decays with neutrino emission.
The dominant decay mode is $p\to K^+\nu_{e}$ and the corresponding
amplitude is proportional to
$\fr{1}{M_G}2\lam_e\lam_c\te_c\ep \ep_1$ ($\te_c=0.22$ is a Cabibbo angle).
Note that in MSSU5 the amplitude of the dominant decay mode
$p\to K\nu_{\mu }$ is$\sim \fr{1}{M_G^0}2\lam_s\lam_c\te_c^2$. Thus,
in our model for
the corresponding partial life time we expect
$$
\tau_{d=5} (p\to K^+\nu_e)=
$$
\beq
\l \fr{\lam_s\te_c}{\lam_e\ep \ep_1}\r^2
\l \fr{M_G}{M_G^0}\r^2 \l \fr{(A^S_{d=5})^0}{A^S_{d=5}}\r^2
\tau_0(p\to K^+\nu_{\mu })~,
\la{model-t-pnu}
\eeq
where $\tau_0$ is proton life time  in MSSU5.  Taking all SUSY breaking soft
terms near TeV scale, we have
$$
\tau_{d=5} (p\to K^+\nu_e)\simeq
$$
\beq
3.8\cdot 10^3\cdot \l \fr{0.1}{\ep}\r^2
\l \fr{1/3}{\ep_1}\r^2
\l \fr{M_G}{2.2\cdot 10^{15}{\rm GeV}}\r^2 \tau_0~.
\la{model-life-pnu}
\eeq
In (\ref{model-life-pnu}) we  used
$A^S_{d=5}=2.03$ calculated for our model [see Eq. (\ref{model-AS-d5})].
The decays with emission of the charged leptons are due to $\al'$
factors. The dominant mode is $p\to K^0\mu^+$ (with corresponding
factor ${\al'}_{22}\simeq 2\lam_u\lam_{\mu }\ep \ep_1$) with the life time
$$
\tau_{d=5} (p\to K^0\mu^+)\simeq
$$
\beq
5.4\cdot 10^3\cdot \l \fr{0.1}{\ep}\r^2
\l \fr{1/3}{\ep_1}\r^2
\l \fr{M_G}{2.2\cdot 10^{15}{\rm GeV}}\r^2 \tau_0~.
\la{model-life-pmu}
\eeq
  As we see, both decay modes of Eqs. (\ref{model-life-pnu}), (\ref{model-life-pmu}) are
suppressed in comparison to the dominant decay mode of MSSU5. In order to make an estimate of proton
life time one should make selection of sparticle spectrum. With soft terms near TeV, 
 given in (\ref{soft-spectrum}),  we will have  \cite{ft4}
$\tau_0\simeq 3.5\cdot 10^{30}$~years \cite{Berezhiani:1998hg}.
Thus, we will have
$$
\tau_{d=5} (p\to K^+\nu_e)\simeq 0.7\cdot  \tau_{d=5} (p\to K^0\mu^+)
\simeq
$$
\beq
1.3\cdot 10^{34}~{\rm years}\tm \l \fr{\sin 2\bt}{0.50}\r^2~.
\la{pred-p}
\eeq
These are above current experimental bounds $\tau^{\rm exp }(p\to K^+\nu )\stackrel{>}{_\sim }6.7\cdot 10^{32}$~years.
and $\tau^{\rm exp }(p\to K^0\mu^+ )\stackrel{>}{_\sim }1.2\cdot 10^{32}$~years \cite{Yao:2006px}. Ongoing and planned experiments give
promise to probe partial lifetimes given in (\ref{pred-p}) (these life times decrease with increase of $\tan \bt $).

Since in our model the GUT scale is reduced nearly by factor $10$ and the unified gauge
coupling is stronger, the $d=6$ operators become relevant. However, doe to multiplet splitting,
the suppression still occurs [see Eq. (\ref{d6ops})]. The dominant $d=6$ operator is
$\ep_1^2\fr{g^2}{M_G^2}(q_1q_1u_1^{c\dagger }e_1^{c\dagger })_D$, where subscripts label the
flavor indices. This operator induces the process $p\to \pi^0e^+$ with a decay
width:
\beq
\Ga_{d=6} (p\to \pi^0e^+)=\fr{m_p}{16\pi f_{\pi }^2}\bar{\al }^2(1+D+F)^2
\l \fr{g^2}{M_G^2}\ep_1^2A^{d=6}_R\r^2~.
\la{Ga-p-epi}
\eeq
With $f_{\pi }=0.13$~GeV, $\bar{\al }=0.015~{\rm GeV}^3$, $D=0.8$, $F=0.47$,
$A^{d=6}_R=2.99$  and $\ep_1=1/3$ we get
\beq
\tau_{d=6}(p\to \pi^0e^+)=\fr{1}{\Ga_{d=6} (p\to \pi^0e^+)}
\simeq 5\cdot 10^{33}~{\rm years}~,
\la{tau-d6}
\eeq
which is slightly above the experimental limit [$\tau^{\rm exp }(p\to \pi^0e^+ )\stackrel{>}{_\sim }1.6\cdot 10^{33}$~yrs.]
This (possibly) dominant decay mode is a characteristic signature of our model.
Future experiments will probe such decays and test viability of the
particular model presented here.

\section{Conclusions}

In this paper we have suggested the mechanism for suppressing the nucleon
decay within SUSY $SU(5)$ GUT. The mechanism is based on idea of split
multiplets and also helps to build realistic fermion pattern. For transparent
demonstration of the presented mechanism we have considered simple example
consistent with gauge coupling unification, realistic fermion mass pattern  and the proton life
time compatible with experiments.

The suggested possibilities can be applied  for building
various realistic $SU(5)$ scenarios with interesting phenomenological
implications. In particular, it would be interesting, in this context, to
address the problem of flavor and try to gain a natural understanding of
observed hierarchies between fermion masses and mixings. Also, it is
desirable to understand the origin of hierarchies between various mass
scales appearing in the construction. For all this an additional symmetries,
such as flavor symmetry, may play crucial role  guaranteeing the robustness of
predictions.
In a concrete model, for realizing suggested mechanisms and for suppression of unwanted baryon number violation,
we have applied ${\cal U}(1)\tm {\cal Z}_{3N}$ symmetry (also providing an 
automatic $R$-parity). It will be interesting to use such a symmetry as a flavor
symmetry.

Finally, here we have not attempted to have natural solution of the doublet-triplet splitting problem.
For the latter  GUTs such as $SO(10)$ \cite{Dimopoulos:1981xm} and
$SU(6)$ \cite{Berezhiani:1989bd}, \cite{Shafi:1999tn} are more motivated.
One can attempt to realize the split multiplet  mechanism
within these constructions and also study  other phenomenology.
These and related issues will be discussed elsewhere.

\vs{0.5cm}

\hs{-0.3cm}{\bf Acknowledgments}

\vs{0.2cm}
\hs{-0.3cm}I thank K.S. Babu for useful discussions.

\bibliography{apssamp}

\begin{thebibliography}{99}
\bibitem{ft2}
Although we will discuss the mechanism in an example of SUSY $SU(5)$, it can be applied
also for non SUSY $SU(5)$ in order to suppress GUT gauge boson mediated $d=6$ baryon
number violation.
\bibitem{ftDT}
Of course,
it is desirable to have more appealing mechanism for the DT splitting, but this goes beyond the scope of this work.
\end{thebibliography}

\bibliographystyle{unsrt}

\end{document}